\documentclass[twocolumn,english,aps,prb,floats]{revtex4}
\usepackage[T1]{fontenc}
\usepackage[latin9]{inputenc}
\usepackage{bm}
\usepackage{amsmath}
\usepackage{amssymb}
\usepackage{graphicx}
\usepackage{esint}
\usepackage{adjustbox}
\usepackage{longtable}
\makeatletter
\@ifundefined{textcolor}{}
{%
 \definecolor{BLACK}{gray}{0}
 \definecolor{WHITE}{gray}{1}
 \definecolor{RED}{rgb}{1,0,0}
 \definecolor{GREEN}{rgb}{0,1,0}
 \definecolor{BLUE}{rgb}{0,0,1}
 \definecolor{CYAN}{cmyk}{1,0,0,0}
 \definecolor{MAGENTA}{cmyk}{0,1,0,0}
 \definecolor{YELLOW}{cmyk}{0,0,1,0}
}

\usepackage{babel}

\makeatother

\usepackage{babel}
\begin{document}

\title{Integral Absorption of Microwave Power by Random-Anisotropy Magnets}
\author{Eugene M. Chudnovsky and Dmitry A. Garanin}
\affiliation{Physics Department, Herbert H. Lehman College and Graduate School,
The City University of New York, 250 Bedford Park Boulevard West,
Bronx, New York 10468-1589, USA }
\date{\today}
\begin{abstract}
We study analytically, within a continuous field model, and numerically on lattices containing $10^5$ spins, the integral absorption of microwaves by a random-anisotropy magnet, $\int d\omega P(\omega)$. It scales as $D_R^2/J$ on the random-anisotropy strength $D_R$ and the strength of the ferromagnetic exchange $J$ in low-anisotropy amorphous magnetic materials. At high anisotropy and in low-anisotropy materials sintered of sufficiently large ferromagnetic grains, the integral power scales linearly on $D_R$. The maximum bandwidth, combined with the maximum absorption power, is achieved when the amorphous structure factor, or grain size, is of an order of the domain wall thickness in a conventional ferromagnet that is of the order of $(J/D_R)^{1/2}$ lattice spacings. 
\end{abstract}
\maketitle

\section{Introduction}
\label{Intro}

Broadband absorption of electromagnetic radiation is a desirable feature for many technological applications \cite{carbon2020}, such as, e.g., microwave shielding, thermal cancer treatment, and stealth technology to name a few. Magnetic materials are commonly used for that purpose. They typically consist of small ferromagnetic particles embedded in a non-magnetic matrix \cite{nanocomposites}. 

If the particles are metallic, their size must be small compared to the skin depth. Otherwise, the absorption of electromagnetic energy would be limited to the skin layer. For a good conductor, the latter is less than one micrometer in depth when the frequency of the radiation is in the upper gigahertz range that is explored, for example, by modern radar technology. At such frequencies, a good absorber using metallic magnetic particles should consist of particles in the nanometer range. 

The fraction of the magnetic volume in the absorbing material poses another limitation on the absorbed radiation power. The particles must be densely packed, which is best achieved in sintered magnets \cite{Cui2022}. Still, if the nanoparticles are metallic, they have to be coated \cite{Evengelista-MMM2021} with a thin insulating layer to make the material nonconducting. This, again, must result in the loss of the magnetic volume absorbing radiation. Thus, an ideal system would be an amorphous or nanocrystalline magnetic material \cite{Martin2020}. Numerous dielectric amorphous ferromagnets have been synthesized in recent years \cite{PSS2016}.

\begin{figure}[h]
\centering{}\includegraphics[width=8cm]{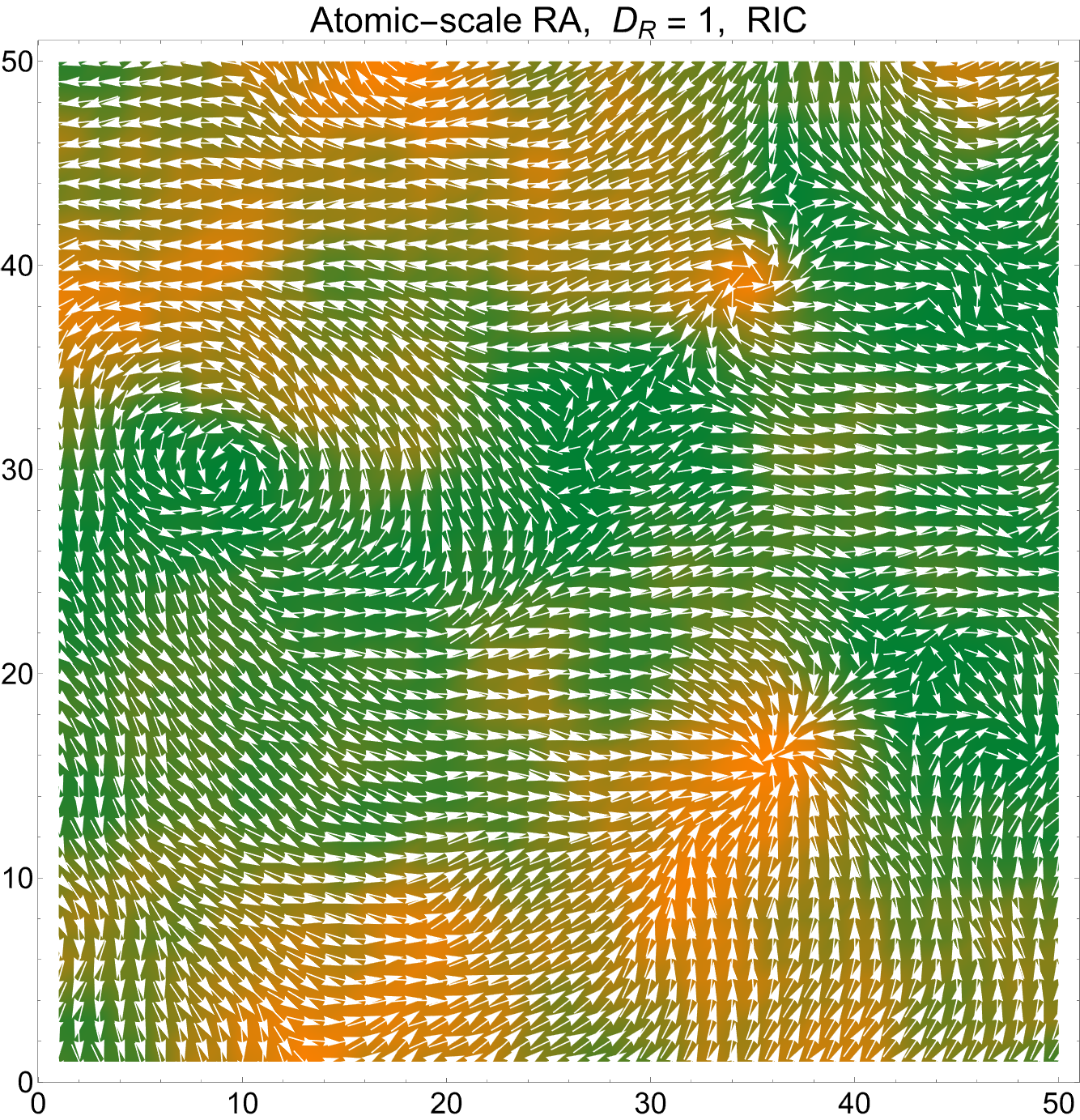}\
\centering{}\includegraphics[width=8cm]{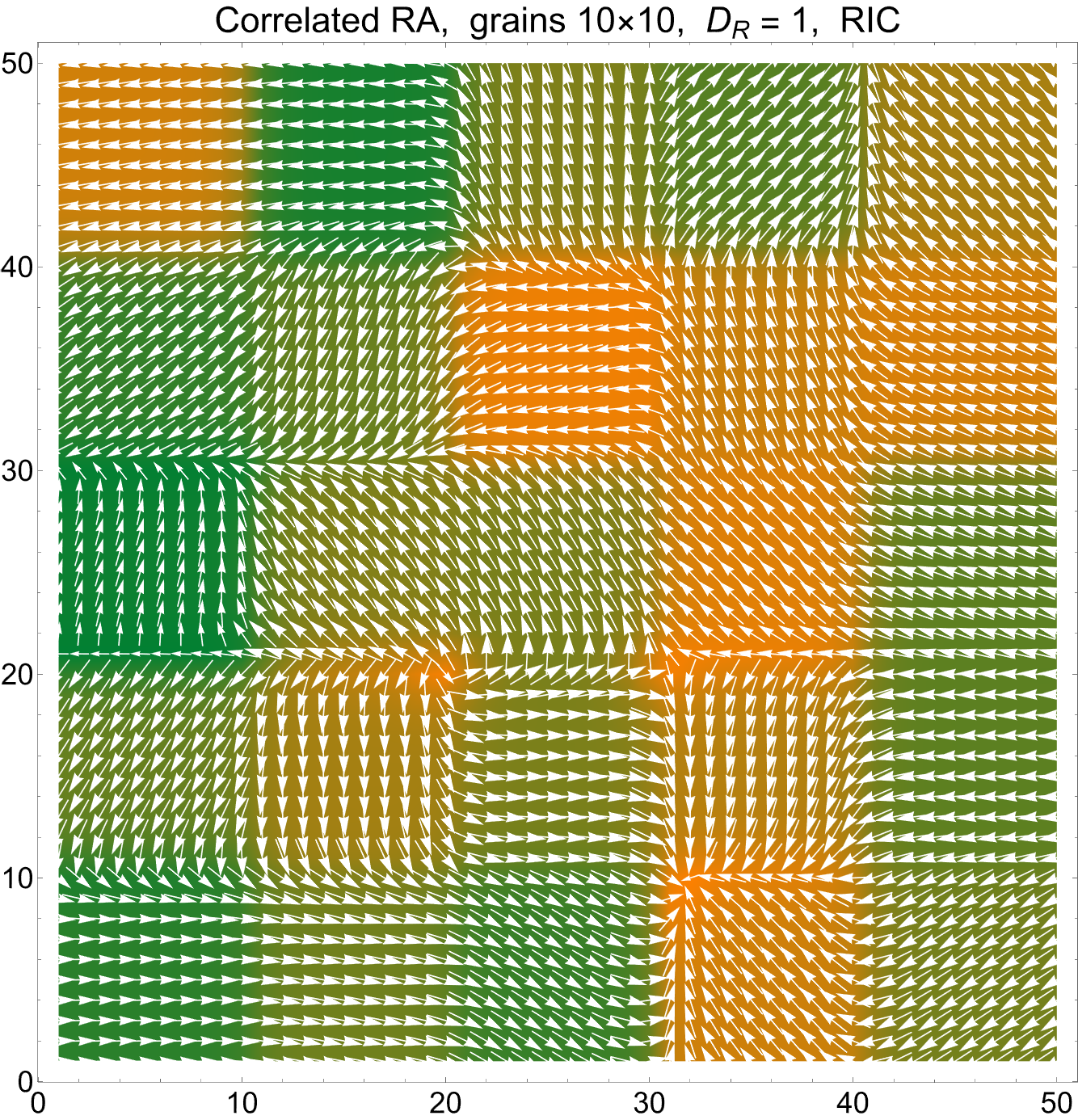}\
\caption{Upper panel: Equilibrium spin structure obtained numerically in a 2D amorphous ferromagnet with random anisotropy axes of individual spins. In-plane spin components are shown by white arrows. The out-of-plane component is shown by orange/green corresponding to positive/negative. Ferromagnetically oriented regions grow on increasing $J/D_R$. Lower panel: Spin structure of a ferromagnet sintered from randomly oriented nanograins having the same anisotropy direction for all spins inside the grain.}
\label{amorphous} 
\end{figure}

Recently, it was shown \cite{GC-PRB2021,GC-PRB2022}  that random-anisotropy amorphous magnets can be promising materials for broadband microwave absorption. In a way, they represent the ultimate limit of densely packed ferromagnetic grains, with no magnetic volume lost, as is illustrated by Fig. \ref{amorphous}. A polycrystalline system consisting of nanoscale ferromagnetic crystallites, or a system sintered of nanoscale ferromagnetic grains, provides a similar advantage. The theoretical study of such systems presents a greater challenge than studies of arrays of weakly interacting ferromagnetic particles due to the exchange interaction between ferromagnetically ordered regions shown in Fig. \ref{amorphous}.

Static properties of amorphous and sintered magnets have been studied theoretically within the random-anisotropy (RA) model for four decades, see, e.g, Refs.\ \onlinecite{RA-book,CT-book,PCG-2015}
and references therein. The model assumes a ferromagnetic exchange of strength $J$ between neighboring spins and random magnetic anisotropy of strength $D_R$ for each spin. The properties of the magnet depend on the ratio $D_R/J$. In practice, magnetic anisotropy arises from relativistic interactions while the exchange comes from the Coulomb between the electrons. Thus, in general, $D_R \ll J$ at the atomic scale. In this case, the magnetic anisotropy cannot win locally over the ferromagnetic exchange and the neighboring spins assume parallel alignment everywhere except around topological defects \cite{PGC-PRL,CG-PRL} seen Fig. \ref{amorphous}. Random pushes of the RA, however, prevent the system from long-range ordering on quenching from the paramagnetic state in a zero magnetic field  \cite{IM,CSS-1986}, resulting in the finite ferromagnetic correlation length that determines the average size of ferromagnetically ordered regions seen in Fig. \ref{amorphous}.

In the limit of very large $D_R$ each spin points in one of the two directions along the local anisotropy axis. This limit is difficult to realize in amorphous ferromagnets with atomic disorder. However, when the amorphous structure factor (correlation length of the structural order) is greater than the atomic spacing, or in the RA magnet sintered of nanocrystallites, it is the magnetic anisotropy energy of the nanocrystallite, proportional to its size, that defines the spin structure. Both the effective anisotropy and the effective exchange depend on the average size of the crystallite or an amorphous structure factor. Consequently all three situations $D_R \ll J$, $D_R \sim J$, and $D_R \gg J$ can be realized in practice \cite{PCG-2015}. High nonlinearity and metastability of the RA problem still evade rigorous results. 

For that reason theoretical studies of the dynamical properties of the RA magnets have been scarce, mainly focused on the ferromagnetic resonance \cite{Saslow2018} (FMR) studied experimentally in random magnets \cite{Monod,Prejean,Alloul1980,Schultz,Gullikson}, and the localization of spin modes \cite{Fert,Levy,Henley1982,HS-1977,Saslow1982,Bruinsma1986,Serota1988,Ma-PRB1986,Zhang-PRB1993,Alvarez-PRL2013,Yu-AnnPhys2013,Nowak2015} that has been reported in various disordered magnetic systems  \cite{Amaral-1993,Suran1-1997,Suran2-1997,Suran-1998,McMichael-PRL2003,Loubens-PRL2007,Du-PRB2014}. More recently, localized spin-wave excitations generated by microwaves in the RA system, as well as their dependence on the RA and the connection between the localization and power absorption, have been investigated numerically \cite{GC-arXiv} on spin lattices containing up to $10^5$ spins. 

In this article, we study the integral microwave power absorption by an RA ferromagnet, defined as $\int d\omega P(\omega)$. While the frequency dependence of the power $P(\omega)$ is difficult to compute analytically, the dependence of the integral power (IP) on parameters can be computed rigorously due to symmetry rules in the limits of weak and large RA. It scales as $D_R^2/J$ in low-anisotropy amorphous magnetic materials. At high anisotropy and in low-anisotropy materials sintered of sufficiently large ferromagnetic grains, the IP scales linearly on $D_R$. Our numerical results agree with the conclusions of the analytical theory. We show that the maximum broadband absorption of the microwave power occurs for the amorphous structure factor, or a grain size of the order of the domain wall thickness of a conventional ferromagnet $(J/D_R)^{1/2}a$, with $a$ being the lattice spacing. 

The paper is organized as follows. The general formula for the IP absorption in a magnetic system is derived in Section \ref{General}. Section \ref{Coherent} provides formulas (needed for comparison with limiting cases of the RA ferromagnet) for the IP in a conventional ferromagnet with uniaxial anisotropy and in a system composed or randomly oriented noninteracting ferromagnetic grains. Integral power absorption in the RA ferromagnet disordered at the atomic scale is derived in Section \ref{RA}. Correlated disorder, when directions of local anisotropy axes are correlated on a scale $R_a > a$, is studied in Section \ref{Correlated}. Numerical results are presented in Section \ref{Numerical}. Our findings are summarized in Section \ref{Discussion}. 

\section{Integral Power Absorption by Magnetic Substance}
\label{General}

We consider a magnet or an array of magnets whose size is small compared to the wavelength of the electromagnetic wave so that the magnet is effectively acted upon by the uniform oscillating magnetic field of the microwave radiation. The power of the radiation of frequency $\omega$ and amplitude $h$, that is absorbed by the magnetic material containing $N$ spins, is proportional to the imaginary part of the susceptibility $\chi''(\omega)$ per spin,
\begin{equation}
P(\omega) = \frac{1}{2}\omega \chi''(\omega)h^2N.
\end{equation}
Fluctuation-dissipation theorem yields
\begin{equation}
\chi''(\omega) =\frac{\omega}{2T}S(\omega), 
\end{equation}
so that
\begin{equation}
P(\omega) = \frac{h^2\omega^2}{4T} S(\omega)N,
\end{equation}
where $T$ is the absolute temperature in energy units and $S(\omega)$ is the time Fourier transform of the correlation function of the total magnetic moment of the system, 
\begin{eqnarray}
S(\omega) & = & \int^{\infty}_{-\infty} dt e^{i\omega(t-t')}\langle {M}_x(t){M}_x(t')\rangle \nonumber \\
& = &\frac{1}{3}\int^{\infty}_{-\infty} dt e^{i\omega(t-t')}\langle {\bf M}(t)\cdot{\bf M}(t')\rangle ,
\end{eqnarray}
For normalization purpose, it is written in terms of the magnetization ${\bf M}$ defined as the magnetic moment per spin of the substance. Here we have assumed that the electromagnetic radiation is polarized along the $x$-axis and that the properties of the magnet are spatially isotropic. Thus,
\begin{eqnarray}
&& \int^{\infty}_{-\infty} \frac{d\omega}{2\pi} P(\omega)  = \\
&& -\frac{h^2N}{12T}\int^{\infty}_{-\infty} \frac{d\omega}{2\pi} \int^{\infty}_{-\infty} dt \left[\frac{d^2}{dt^2}e^{i\omega(t-t')}\right]\langle {\bf M}(t)\cdot{\bf M}(t')\rangle = \nonumber \\
 && -\frac{h^2N}{12T}\int^{\infty}_{-\infty} \frac{d\omega}{2\pi} \int^{\infty}_{-\infty} dt e^{i\omega(t-t')}\frac{d^2}{dt^2}\langle {\bf M}(t)\cdot{\bf M}(t')\rangle = \nonumber \\
 && \frac{h^2N}{12T}\int^{\infty}_{-\infty} \frac{d\omega}{2\pi} \int^{\infty}_{-\infty} dt e^{i\omega(t-t')}\frac{d}{dt}\frac{d}{dt'}\langle {\bf M}(t)\cdot{\bf M}(t')\rangle, \nonumber
 \end{eqnarray}
 where we have used the fact that $\langle {\bf M}(t)\cdot{\bf M}(t')\rangle$ depends on $t-t'$.
 Finally, one obtains for the IP in the volume $V$
 \begin{eqnarray}
&& \int^{\infty}_{-\infty} \frac{d\omega}{2\pi} P(\omega) =  \\ \label{IP-DFT}
&& -\frac{h^2N}{12T}\int^{\infty}_{-\infty} dt \int^{\infty}_{-\infty} \frac{d\omega}{2\pi} e^{i\omega(t-t')} \langle \dot{\bf M}(t)\cdot \dot{\bf M}(t')\rangle = \nonumber \\
& & \frac{h^2N}{12T}\int^{\infty}_{-\infty} dt \delta(t-t')\langle \dot{\bf M}(t)\cdot \dot{\bf M}(t')\rangle = \frac{h^2N}{12T}\langle \dot{\bf M}^2(t)\rangle, \nonumber
\end{eqnarray}
where $\dot{\bf M} \equiv d{\bf M}/dt$. This formula is similar to the sum rules in quantum field theory. 

In what follows we write the magnetization as ${\bf M} = \mu\sum_{\bf r}{\bf s}_{\bf r} /N$, where $\mu$ is an individual magnetic moment and $\sum_{\bf r}{\bf s}_{\bf r} $ is the sum over $N$ spins. Then, expressing $h$ in the energy units via the Zeeman formula, $h \rightarrow \mu h$ one can re-write the formula for the integral power absorption by $N$ spins as 
\begin{equation}
 \int^{\infty}_{-\infty} \frac{d\omega}{2\pi} P(\omega) = \frac{h^2}{12NT}\Big\langle \left(\frac{d}{dt}\sum_{\bf r}{\bf s}_{\bf r}\right)^2\Big\rangle.
 \label{IP-general-N}
 \end{equation}

\section{Integral Power in a Ferromagnet with Coherent Magnetic Anisotropy}
\label{Coherent}

For reference, we will begin with obtaining the integral power (IP) for the textbook problem of ferromagnetic resonance. The expression for the power absorption in the presence of the dimensionless damping $\eta$, no stationary external magnetic field, and uniaxial anisotropy with the energy constant $D$ is given by (see, e.g., Ref.\ \onlinecite{CT-book}) 
 \begin{equation}
 P(\omega) =  \frac{DN({\bf n} \times {\bf h})^2}{2\hbar^2}\frac{2\eta \omega_{\rm FMR}^3}{(\omega^2 - \omega_{\rm FMR}^2)^2 -( 2\eta \omega_{\rm FMR}^2)^2},
 \label{P-FMR}
 \end{equation}
where $ \omega_{\rm FMR} = D/\hbar $ is the frequency of the ferromagnetic resonance and ${\bf n}$ is the unit vector along the anisotropy axis. Here we neglected magnetic dipolar fields (considering them small compared to the anisotropy energy) and expressed the ac field ${\bf h}$ acting on individual spins in the energy units. 
 
We first consider one crystallite in the ac field perpendicular to the anisotropy axis, when $({\bf n} \times {\bf h})^2 = h^2$. In the limit of vanishing damping, $\eta \rightarrow 0$, the second fraction in Eq.\ (\ref{P-FMR}) becomes the $\delta$-function,
\begin{eqnarray}
&& P(\omega) = \frac{Dh^2N}{2\hbar^2}\omega_{\rm FMR}\pi\delta({\omega}^2 - {\omega}_{\rm FMR}^2) \nonumber \\
&& = \frac{\pi Dh^2N}{4\hbar^2}[\delta({\omega} - {\omega}_{\rm FMR}) + \delta({\omega} + {\omega}_{\rm FMR})],
\end{eqnarray}
and the IP acquires a universal form:
\begin{equation}
 \int^{\infty}_{-\infty} \frac{d\omega}{2\pi} P(\omega)   = \frac{Dh^2}{4\hbar^2}N.
 \label{IP-coherent}
 \end{equation}
The power absorption is sharply peaked at $\omega = {\omega}_{\rm FMR}$. 

Eq.\ (\ref{IP-coherent}) can be easily generalized for the case of randomly oriented noninteracting ferromagnetic grains by averaging $({\bf n} \times {\bf h})^2$ over the random directions of the anisotropy axis ${\bf n}$ in Eq (\ref{P-FMR}). As a result, Eq.\ (\ref{IP-coherent}) becomes multiplied by a factor
 \begin{equation}
 \frac{1}{4\pi}\int_0^{2\pi} d \phi \int_0^{\pi} \sin\theta d \theta \, sin^2{\theta} = \frac{2}{3}
  \end{equation}
 This gives for the IP
  \begin{equation}
 \int^{\infty}_{-\infty} \frac{d\omega}{2\pi} P(\omega)  =  \frac{D_R h^2}{6\hbar^2}N.
 \label{IP-distributed}
 \end{equation}
Even in this case, however, one needs distribution over FMR frequencies to make the power absorption broadband. In practice, this is achieved due to the variation in the shape of the particles and the resulting variation in their surface anisotropy. 
  
 \section{Integral Power in a Random Anisotropy Ferromagnet}
 \label{RA}
 
The simplest way to increase the strength of the broadband power absorption for a given magnetic material is to increase the fraction of the magnetic volume inside the sample. This can be achieved by compressing the powder of dielectric ferromagnetic particles or densely packing metallic particles that are small compared to the thickness of the skin layer, coated with a thin insulating layer. The ultimate limit is provided by the material sintered of nanoscale ferromagnetic particles or, even better, by a nonconducting amorphous ferromagnetic material. This limit, however, is more challenging for theoretical studies because it must include the exchange interaction between the particles in a sintered ferromagnet or between individual spins in an amorphous ferromagnet. We shall start with the latter problem in which the disorder in the orientation of anisotropy axes occurs at the smallest scale of individual spins.

To compute the sum in the right-hand-side of Eq.\ (\ref{IP-general-N}) we shall use the discrete version of the random-anisotropy problem described by the Hamiltonian
\begin{equation}
{\cal{H}} = -\frac{J}{2}\sum_{{\bf r},{\bf r}'}{\bf s}_{\bf r} \cdot {\bf s}_{{\bf r}'} - \frac{D_R}{2}\sum_{\bf r}({\bf n}_{\bf r} \cdot {\bf s}_{\bf r})^2,
\label{Hamiltonian}
\end{equation}
where ${\bf s}_{\bf r}$ and ${\bf n}_{\bf r}$ are the unit spin vector and the unit vector of local magnetic anisotropy at the position given by the radius vector ${\bf r}$ respectively, $J$ and $D_R$ are the ferromagnetic exchange and anisotropy strength constants, and summation in the exchange term is over the nearest neighbors with ${{\bf r} \neq {\bf r}'}$. The equation of motion for each spin is
\begin{equation}
\hbar \frac{d{\bf s}_{\bf r}}{dt} = {\bf s}_{\bf r} \times \left[J\sum_{{\bf r}' \neq {\bf r}}{\bf s}_{{\bf r}'}   + D_R({\bf n}_{\bf r} \cdot {\bf s}_{\bf r}){\bf n}_{\bf r} \right],
\label{LL}
\end{equation}
which yields
\begin{equation}
\frac{d}{dt} \sum_{\bf r}{\bf s}_{{\bf r}}= \frac{D_R}{\hbar}\sum_{\bf r} ({\bf s}_{\bf r} \cdot {\bf n}_{\bf r})({\bf s}_{\bf r} \times {\bf n}_{\bf r}).
\label{IP-numerical}
\end{equation}
The first term in Eq.\ (\ref{LL}) vanishes by symmetry. 

We now notice that the time dependence of spins is generated by their deviation from equilibrium static orientations, ${\bf s}_{\bf r} = {\bf s}^{(0)}_{\bf r} + \delta {\bf s}_{\bf r}$. This gives
\begin{eqnarray}
&& \frac{d}{dt} \sum_{\bf r}{\bf s}_{{\bf r}} = \\
&& \frac{D_R}{\hbar}\sum_{\bf r} \left[({\bf s}^{(0)}_{\bf r} \cdot {\bf n}_{\bf r})(\delta{\bf s}_{\bf r} \times {\bf n}_{\bf r}) + (\delta{\bf s}_{\bf r} \cdot {\bf n}_{\bf r})({\bf s}^{(0)}_{\bf r} \times {\bf n}_{\bf r})\right]. \nonumber
\end{eqnarray}
In what follows we shall assume that the ferromagnetic correlation length $R_f$ (which in accepted terms represents the average size of the IM domain in the RA ferromagnet) is large compared to the distance between neighboring spins $a$. This allows one to write Eq.\ (\ref{IP-general-N}) for the IP as 
\begin{eqnarray}
 && \int^{\infty}_{-\infty} \frac{d\omega}{2\pi} P(\omega)  = \frac{D_R^2h^2}{12\hbar^2NT} \times \nonumber \\
 && \sum_{{\bf r},{\bf r}'}  \Big\langle \left[\epsilon_{ijk}s^{(0)}_i({\bf r})n_j({\bf r})\delta s_m({\bf r})n_m({\bf r}) +  \right. \nonumber \\
 && \left. s_m^{(0)}({\bf r})n_m({\bf r}) \epsilon_{ijk} \delta s_i({\bf r}) n_j({\bf r}) \right] \times \nonumber \\
 && \left[\epsilon_{lnk}s^{(0)}_l({\bf r}')n_n({\bf r}')\delta s_r({\bf r}')n_r({\bf r}') + \right. \nonumber \\
 && \left. s_r^{(0)}({\bf r}')n_r({\bf r}') \epsilon_{lnk} \delta s_l({\bf r}') n_n({\bf r}') \right]\Big\rangle = \nonumber \\
&& \frac{D_R^2h^2}{12\hbar^2NT}\sum_{{\bf r},{\bf r}'}\epsilon_{ijk}\epsilon_{lnk}\langle n_j({\bf r})n_m({\bf r})n_r({\bf r}')n_n({\bf r}')\rangle \times \nonumber \\
&& \Big\langle \left[s^{(0)}_i({\bf r})\delta s_m({\bf r})+ s_m^0({\bf r})\delta s_i({\bf r}) \right] \times \nonumber \\
&& \left[s^{(0)}_l({\bf r}')\delta s_r({\bf r}') + s_r^{(0)}({\bf r}') \delta s_l({\bf r}')\right]\Big\rangle,
\label{RAsum}
\end{eqnarray}
with lower indices of variables corresponding to three components of ${\bf s}$ and ${\bf n}$. We have used random vector ${\bf n}$ and the fact that local directions of magnetization and anisotropy axes are very weakly correlated when $R_f \gg a$. 

For the fourth moment of the Gaussian distribution of anisotropy axes, we write
\begin{eqnarray}
&&\langle n_j({\bf r})n_m({\bf r})n_r({\bf r}')n_n({\bf r}')\rangle = \nonumber \\
&& \langle n_j({\bf r})n_m({\bf r})\rangle\langle n_r({\bf r}')n_n({\bf r}')\rangle + \nonumber \\
&& \langle n_j({\bf r})n_r({\bf r}')\rangle \langle n_m({\bf r})n_n({\bf r}')\rangle + \nonumber \\
&& \langle n_j({\bf r})n_n({\bf r}')\rangle \langle n_m({\bf r})n_r({\bf r}')\rangle  \nonumber \\
&& = \frac{1}{3} \delta_{jm}\frac{1}{3}\delta_{rn} + \frac{1}{3}\delta_{jr}\delta_{{\bf r}{\bf r}'}\frac{1}{3}\delta_{mn}\delta_{{\bf r}{\bf r}'} +\frac{1}{3}\delta_{jn}\delta_{{\bf r}{\bf r}'}\frac{1}{3}\delta_{mr}\delta_{{\bf r}{\bf r}'} \nonumber \\
&& =  \frac{1}{9} \left(\delta_{jm}\delta_{rn} + \delta_{jr}\delta_{mn}\delta_{{\bf r}{\bf r}'} +\delta_{jn}\delta_{mr}\delta_{{\bf r}{\bf r}'} \right).
\label{Kron}
\end{eqnarray}
The first term in the last sum of the product of Kronecker symbols does not contribute to the sum in Eq.\ (\ref{RAsum}) because it reduces to the product of two independent sums each containing $\delta {\bf s}$. Thus, Eq.\ (\ref{RAsum}) reduces to
\begin{eqnarray}
 && \int^{\infty}_{-\infty} \frac{d\omega}{2\pi} P(\omega)  = \frac{D_R^2h^2}{108\hbar^2NT} \times \nonumber \\
 && \sum_{{\bf r},{\bf r}'}\left(\delta_{il}\delta_{jn} - \delta_{in}\delta_{jl}\right) \left(\delta_{jr}\delta_{mn}\delta_{{\bf r}{\bf r}'} +\delta_{jn}\delta_{mr}\delta_{{\bf r}{\bf r}'} \right) \times \nonumber \\
 && \Big\langle \left[s^{(0)}_i({\bf r})\delta s_m({\bf r})+ s_m^0({\bf r})\delta s_i({\bf r}) \right] \times \nonumber \\
 && \left[s^{(0)}_l({\bf r}')\delta s_r({\bf r}') + s_r^{(0)}({\bf r}') \delta s_l({\bf r}')\right]\Big\rangle  \nonumber \\
 &&= \frac{D_R^2h^2}{108\hbar^2T}\sum_{\bf r}\left(3\delta_{il}\delta_{rm} - \delta_{im}\delta_{rl}\right) \times \nonumber \\
 && \Big\langle \left[s^{(0)}_i({\bf r})\delta s_m({\bf r})+ s_m^0({\bf r})\delta s_i({\bf r}) \right] \times \nonumber \\
 && \left[s^{(0)}_l({\bf r})\delta s_r({\bf r}) + s_r^{(0)}({\bf r}) \delta s_l({\bf r})\right]\Big\rangle  = \nonumber \\
&& \frac{D_R^2h^2}{108\hbar^2T}\sum_{\bf r}\Big\langle 3 \left[s^{(0)}_i({\bf r})\delta s_m({\bf r})+ s_m^{(0)}({\bf r})\delta s_i({\bf r}) \right] \times \nonumber \\
&& \left[s^{(0)}_i({\bf r})\delta s_m({\bf r}) + s_m^{(0)}({\bf r}) \delta s_i({\bf r})\right] - \nonumber \\
&& \left[s^{(0)}_i({\bf r})\delta s_i({\bf r})+ s_i^{(0)}({\bf r})\delta s_i({\bf r}) \right] \times \nonumber \\
&& \left[s^{(0)}_l({\bf r})\delta s_l({\bf r}) + s_l^{(0)}({\bf r}) \delta s_l({\bf r})\right]\Big\rangle \nonumber \\
&& = \frac{D_R^2h^2}{108\hbar^2T}\sum_{\bf r} \Big\langle\left\{6\left[\delta{\bf s}({\bf r})\right]^2 - 2 \left[{\bf s}^{(0)}({\bf r})\cdot \delta{\bf s}({\bf r})\right]^2\right\}\Big\rangle \nonumber \\
&& = \frac{D_R^2h^2}{108\hbar^2T}\sum_{\bf r} \Big\langle 6\left[\delta{\bf s}({\bf r})\right]^2 - \frac{2}{3}\left[\delta{\bf s}({\bf r})\right]^2\Big\rangle \nonumber \\
&& = \frac{4D_R^2h^2}{81\hbar^2T}\sum_{\bf r} \langle\delta{\bf s}({\bf r})\rangle^2. 
\label{P-deltaS}
\end{eqnarray}

At this point we shall assume that at $R_f \gg a$ spin fluctuations are dominated by the ferromagnetically ordered regions and will use the standard formula for the Heisenberg ferromagnet with three spin components in $d$ dimensions:
\begin{equation}
\langle\delta{\bf s}\rangle^2 = \frac{2T}{J}a^{d-2}\int_{1/(R_f)}^{1/a}\frac{d^d k}{(2\pi)^d}\frac{1}{k^2}.
\label{int}
\end{equation}
With the sum in Eq.\  (\ref{P-deltaS}) providing the factor $N$, this gives 
\begin{equation}
\int^{\infty}_{-\infty} \frac{d\omega}{2\pi} P(\omega) = N\left[\frac{4\ln(C R_f/a)}{81\pi}\right] \frac{D_R^2}{\hbar^2J} h^2
\label{IP-2D}
\end{equation}
in two dimensions, where $C$ is a constant of order unity that we introduced to account for the crudeness of the approximation used for the limits of integration in Eq.\ (\ref{int}). Logarithmic dependence of the power on $R_f$, is a consequence of diverging spin fluctuations in 2D.  

In three dimensions, $\langle\delta{\bf s}\rangle^2$ is dominated by the upper limit of integration in Eq. (\ref{int}). If there is a known short-range order in the lattice structure, this integral can be computed exactly.  For $R_f \gg a$ and spins forming a simple cubic lattice at short distances, one has
\begin{equation}
\int^{\infty}_{-\infty} \frac{d\omega}{2\pi} P(\omega) = N\left(\frac{4W}{3^7}\right) \frac{D_R^2}{\hbar^2J} h^2,
\label{IP-3D}
\end{equation}
where $W = 1.516$ is the Watson integral. Up to a numerical factor of order unity, this answer in 3D must be valid for any other short-range arrangement of spins, including random arrangement in an amorphous ferromagnet disordered at the atomic scale.

While the constant $C$ may be difficult to obtain analytically, the $D_R^2/J$ dominant scaling of the IP in the IM state with $R_f \gg a$ in both 2D and 3D is a robust result. The logarithmic factor in Eq.\ (\ref{IP-2D})  provides an additional logarithmic dependence of the IP on RA through $R_f \propto J/D_R$ in 2D. 

\section{Correlated Disorder}
\label{Correlated} 

For an RA magnet with a short-range order in the orientation of local crystallographic axes that extends beyond the distance between neighboring spins, one should consider magnetic anisotropy that is correlated within regions of size $R_a > a$. Such a model can be rescaled to the original model described by the Hamiltonian (\ref{Hamiltonian}) by considering its continuous counterpart, 
\begin{equation}
{\cal{H} } = \frac{1}{2}\int \frac{d^d r}{a^d}\left[Ja^2(\nabla {\bf s})^2 - D_R({\bf n} \cdot {\bf s})^2\right],
\end{equation}
where ${\bf s}$ is the dimensionless spin-field density. Preserving that spin density, this expression can be rescaled by writing the Hamiltonian in terms of the rescaled exchange $J'$ and anisotropy $D'_R$ acting on the blocks of spins of size $R_a$,
\begin{equation}
{\cal{H}} = \frac{1}{2}\int \frac{d^d r}{R_a^d}\left[J'R_a^2(\nabla {\bf s})^2 - D'_R({\bf n} \cdot {\bf s})^2\right] .
\end{equation}
Comparing it with the expression for $R_a = a$, we obtain
\begin{equation}
J' = J\left(\frac{R_a}{a}\right)^{d-2}, \qquad D'_R = D_R\left(\frac{R_a}{a}\right)^d
\label{scaling}
\end{equation}

Let us check first what the correlated disorder does to the ferromagnetic correlation length. For the atomic disorder defined by $R_a = a$, the latter is given by the Imry-Ma formula
\begin{equation}
\frac{R_f}{a} = k_d \left(\frac{J}{D_R}\right)^{2/(4-d)}, 
\label{IM-Rf}
\end{equation}
where $k_d$ is a numerical factor never computed analytically that depends on the dimensionality of space $d$. For the blocks of spins of size $R_a$, one must have
\begin{equation}
\frac{R_f}{R_a} = k_d \left(\frac{J'}{D'_R}\right)^{2/(4-d)} = k_d \left(\frac{J}{D_R}\right)^{2/(4-d)}\left(\frac{a}{R_a}\right)^{4/(4-d)}
\label{Rf-Ra}
\end{equation} 
which is equivalent to \cite{CSS-1986,CT-book,PCG-2015}
\begin{equation}
\frac{R_f}{a} = k_d \left(\frac{J}{D_R}\right)^{2/(4-d)}\left(\frac{a}{R_a}\right)^{d/(4-d)}.
\label{Ra}
\end{equation}
It suggests that the FM correlation length goes down as $1/R_a^3$ in 3D and as $1/R_a$ in 2D. One should keep in mind, however, that $R_f$ cannot go below $a$. This establishes the upper limit on the size of the correlated spin block that is consistent with the IM formula:
\begin{equation}
\frac{R_a}{a} < k_3^{1/3}\left(\frac{J}{D_R}\right)^{2/3} = \left(\frac{R_f}{a}\right)^{1/3}_{R_a = a}
\end{equation} 
in 3D and
\begin{equation}
\frac{R_a}{a} < k_2\left(\frac{J}{D_R}\right) = \left(\frac{R_f}{a}\right)_{R_a = a}
\end{equation} 
in 2D. If this condition is not satisfied, one has $R_f = R_a$, that is, the system becomes equivalent to an array of densely packed, weakly-interacting, single-domain ferromagnetic particles of size $R_a$. 

The integral power derived in the preceding section under the condition $R_f \gg a$ becomes proportional to
\begin{equation}
\frac{{D'_R}^2}{J'} N' = \frac{{D_R}^2}{J} N \left(\frac{R_a}{a}\right)^2,
\label{scaling-prime}
\end{equation}
where we used Eq.\ (\ref{scaling}) and $N' = N(a/R_a)^{d}$ for the number of correlated blocks in a system of a fixed volume (that is, of a fixed number of spins $N$). This implies that 
up to a logarithm the IP is proportional to $R_a^2$ regardless of the dimensionality of the system $d$, which is a rather strong dependence on the size of the correlated block $R_a$. One should keep in mind, however, that it only applies to blocks of sufficiently small size satisfying $R_a \lesssim R_f$. 

At small $D_R$ and $R_a \lesssim R_f$, Eqs.\ (\ref{IP-2D}), (\ref{Rf-Ra}), and (\ref{scaling-prime})  yield in 2D
\begin{equation}
\int^{\infty}_{-\infty} \frac{d\omega}{2\pi} P(\omega) = N\frac{4D_R^2h^2}{81\pi\hbar^2J}\left(\frac{R_a}{a}\right)^2\ln\left[C\left(\frac{J}{D_R}\right)\left(\frac{a}{R_a}\right)\right]. 
\label{IP-2D-Ra}
\end{equation}
As we shall see, this formula provides a good fit to the numerical data on spin lattices reported in the following section.

As $R_a$ approaches $R_f$, the anisotropy $D'_R$ approaches the rescaled exchange $J'$, see Eq.\ (\ref{Rf-Ra}), and Eq.\ (\ref{scaling-prime}) gives
\begin{equation}
\frac{{D'_R}^2}{J'} N' = D'_R N' = (D_R R_a^d)\left(\frac{N}{R_a^d}\right) = D_R N,
\end{equation} 
as it must be for the FMR absorption of the MW power by independent ferromagnetic particles with coherent anisotropy $D = D_R$, see Eq.\ (\ref{IP-distributed}). Notice that in this limit it does not depend on $R_a$. 

At greater  $R_a$ the response of an individual spin block to the magnetic field is dominated by the magnetic anisotropy of the block. Consequently, the IP for a system consisting of randomly oriented large ferromagnetic grains must be given by Eq.\ (\ref{IP-distributed}). In this limit, within our simplified model assuming a fixed RA strength, the power absorption is no longer broadband. It peaks sharply at a single FMR frequency. In practice, however, the variation in surface anisotropy resulting from different shapes of the grains, as well as domain walls trapped by the grain boundaries, must provide a significant bandwidth even at weak damping. 

\section{Numerical Results}
\label{Numerical}
When studying the power absorption numerically, the most interesting nontrivial case is the IM state of an amorphous or sintered ferromagnet with a large $R_f$, when both magnetic anisotropy and ferromagnetic exchange contribute to the absorption. In this case, one should simulate a system of a size that is large compared to the $R_f$. Thus the main parameter range of interest is $D_R \ll J$. According to Eq.\ (\ref{IM-Rf}) the ferromagnetic correlation length scales as $J/D_R$ in 2D and $(J/D_R)^2$ in 3D. Consequently, the number of spins in a simulated system must be large compared to $(J/D_R)^2$ in 2D and large compared to $(J/D_R)^6$ in 3D. The 3D condition is difficult to satisfy even with modern computational capabilities. In what follows, we stick to a 2D case that describes a thin amorphous film or a material sintered of nanofoils. Notice that the low dimensionality of the magnetic system is beneficial for the absorption of microwave power \cite{GC-PRB2021}. 

The integral absorbed power, Eq.\ (\ref{IP-general-N}), has been numerically computed using
Eq. (\ref{IP-numerical}) for the time derivative of the total spin. It was averaged with the help of standard Metropolis Monte
Carlo arguments for systems of classical spins. At temperature $T$,
spins are successively rotated by a random angle in random directions
and the corresponding energy change $\Delta E$ is computed. The rotation
is accepted if $\exp\left(-\Delta E/T\right)>\mathrm{rand}$, where 
rand is a number in the interval $(0,1)$. The average rotation
angle increases with $T$ and is adjusted in such a way that almost
50\% of all trial rotations get accepted. 

To better explore the phase space of the system and to speed up the thermalization,
it is advantageous to combine Monte Carlo with overrelaxation involving spin rotations by $180^\circ$ around their effective fields.
For the interaction of spins from different sites, such as the exchange,
the overrelaxation conserves energy. In the presence of single-site interactions such as the RA, the overrelaxation does not conserve energy and
cannot be used in its simplest form. To deal with this problem, we applied the thermalized
overrelaxation method proposed in Ref.\ \onlinecite{GC-JPhys2022}. 

We studied the thermalization from the state of randomly oriented spins,
that is, with the random initial condition (RIC), and also from the state with
the collinear initial condition (CIC) that has all spins initially aligned in one direction. The first corresponds to the rapid quenching of the amorphous magnet from the melt or to the sintering the RA magnet from randomly oriented grains in a zero magnetic field, while the second implies doing it in the presence of a strong field. At low temperatures, the RIC yields
a nearly disordered state in which the average spin $\langle\sum_{\mathbf{r}}\mathbf{s}_{\mathbf{r}}\rangle/N$
is small. For the CIC, the disordering due to the RA
is incomplete and there is a significant average spin of about $0.7$ in the final state. 
Still, the microwave power absorption is comparable in both cases. Most of the computations were performed with the CIC.

After the energy of the system and the magnitude of the average spin
reached stationary values, we continued Monte Carlo simulation
to collect the data for the average in the right-hand-side of Eq. (\ref{IP-numerical}). No self-averaging was observed for a large system in accordance with the expectation that quenched randomness results in diverging fluctuations at large distances. Averaging of the statistical scatter was achieved with a long simulation and data collection.
We performed 3000 blocks of 10 Monte Carlo system updates. The values of the right-hand side of Eq. (\ref{IP-numerical}) were recorded at the end of each block.
This provided reasonably smooth curves for the dependences of
the integral absorbed power on $D_{R}$. The system size in most computations was $300\times300$, i.e., nearly $10^{5}$ spins. A long
computation for the $600\times600$ system gave essentially the same
results.

For the model with the atomic-scale RA, we have chosen randomly oriented
anisotropy axes on each lattice site and the same value of $D_{R}$
at each site. For the model with correlated RA, we defined randomly
directed anisotropy axes for spin blocks (grains) of size $2\times2$, $3\times3$,
etc, the anisotropy being the same for all spins within the grain. Computations
were performed in the range of the weak RA, $D_{R}/J\leq0.05$, and
in the range of strong RA, $D_{R}/J\leq2$. 

\begin{figure}[h]
\centering{}\includegraphics[width=9cm]{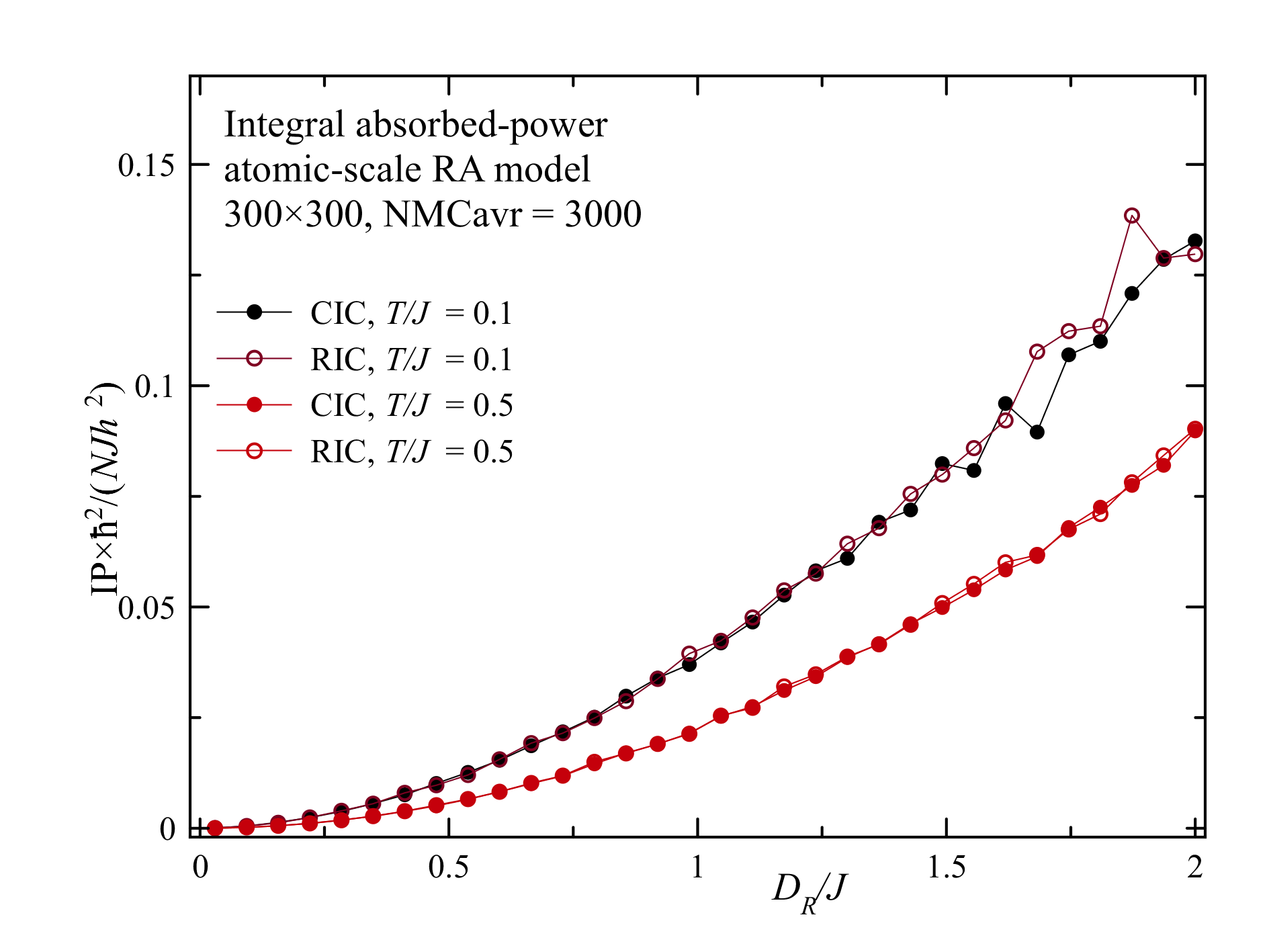}
\caption{The effect of temperature on the integral power absorption by an amorphous magnet disordered at the atomic scale. The data are practically independent of the initial condition, random (RIC) or collinear (CIC).}
\label{T-dependence} 
\end{figure}

Earlier we demonstrated \cite{GC-PRB2022} that elevated temperatures diminish power absorption by the RA magnet due to the increased occupation of the excited spin states. This also applies to the integral power as is illustrated by Fig.\ \ref{T-dependence}. While the effect of heating is strong, the dependence of the IP on the initial spin structure of the magnet, random (RIC) or collinear (CIC), is practically absent. 

\begin{figure}[h]
\centering{}\includegraphics[width=9cm]{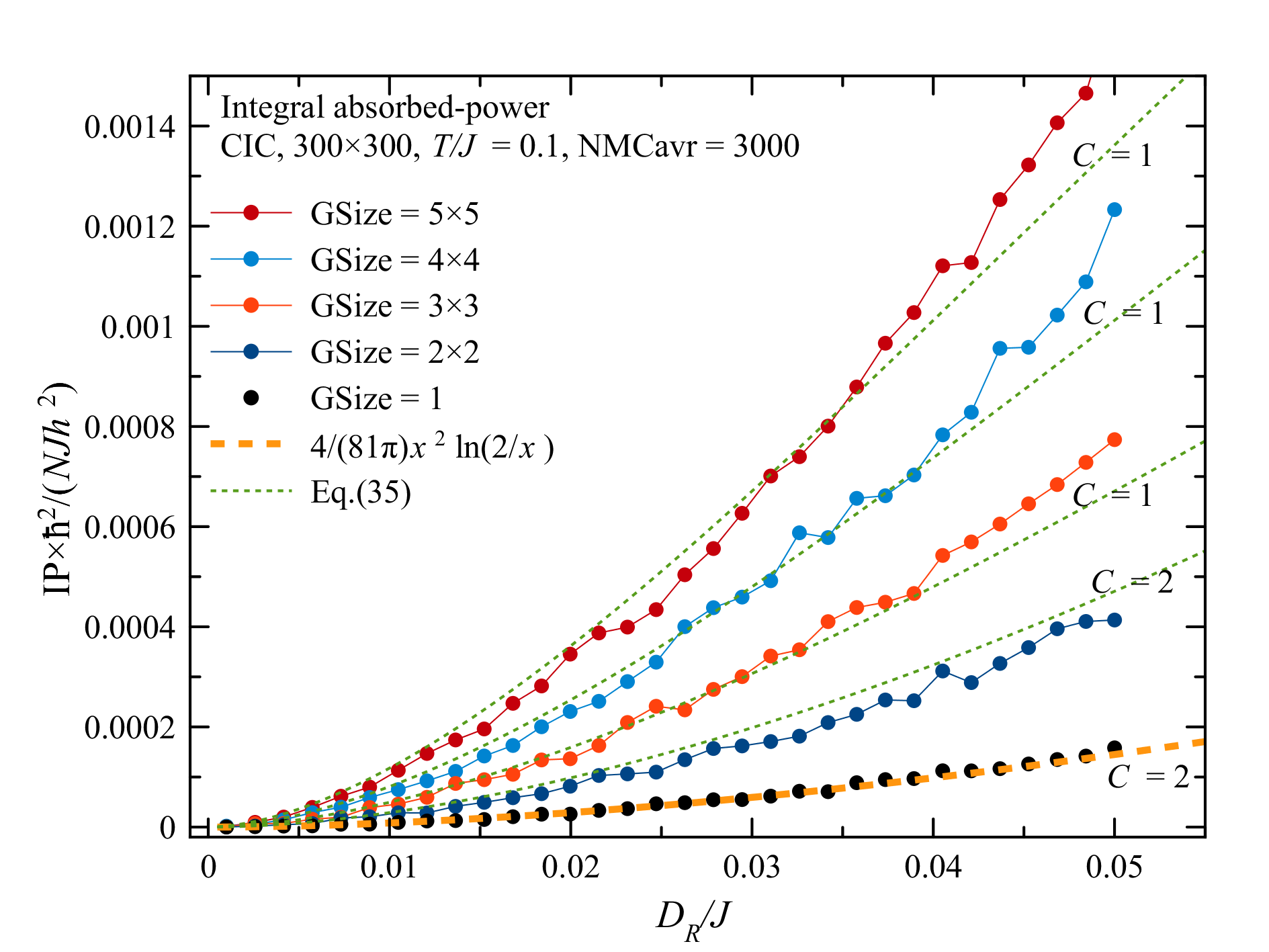}
\caption{Integral power absorption at weak RA computed numerically for the correlated disorder with different grain sizes, GSize =1 corresponding to the disorder in the orientation of the RA axes at the atomic scale. Green dash lines show fit by Eq.\ (\ref{IP-2D-Ra}) in which the GSize is used instead of $R_a^2$. The fitting constant $C$ under the logarithm is indicated next to each curve.}
\label{smallDR} 
\end{figure}
The numerically computed dependence of the IP on $D_R \ll J$ in a 2D system at $T = 0.1J$ is shown for different grain sizes (GSize) in Fig.\ \ref{smallDR}. The data agree with the expectation that the IP increases with the size of the grain. The theoretical formula (\ref{IP-2D-Ra}), with GSize for $R_a^2$ provides a good fit to the numerical data if one adjusts the constant $C$ when the grain size goes up. 

\begin{figure}[h]
\centering{}\includegraphics[width=9cm]{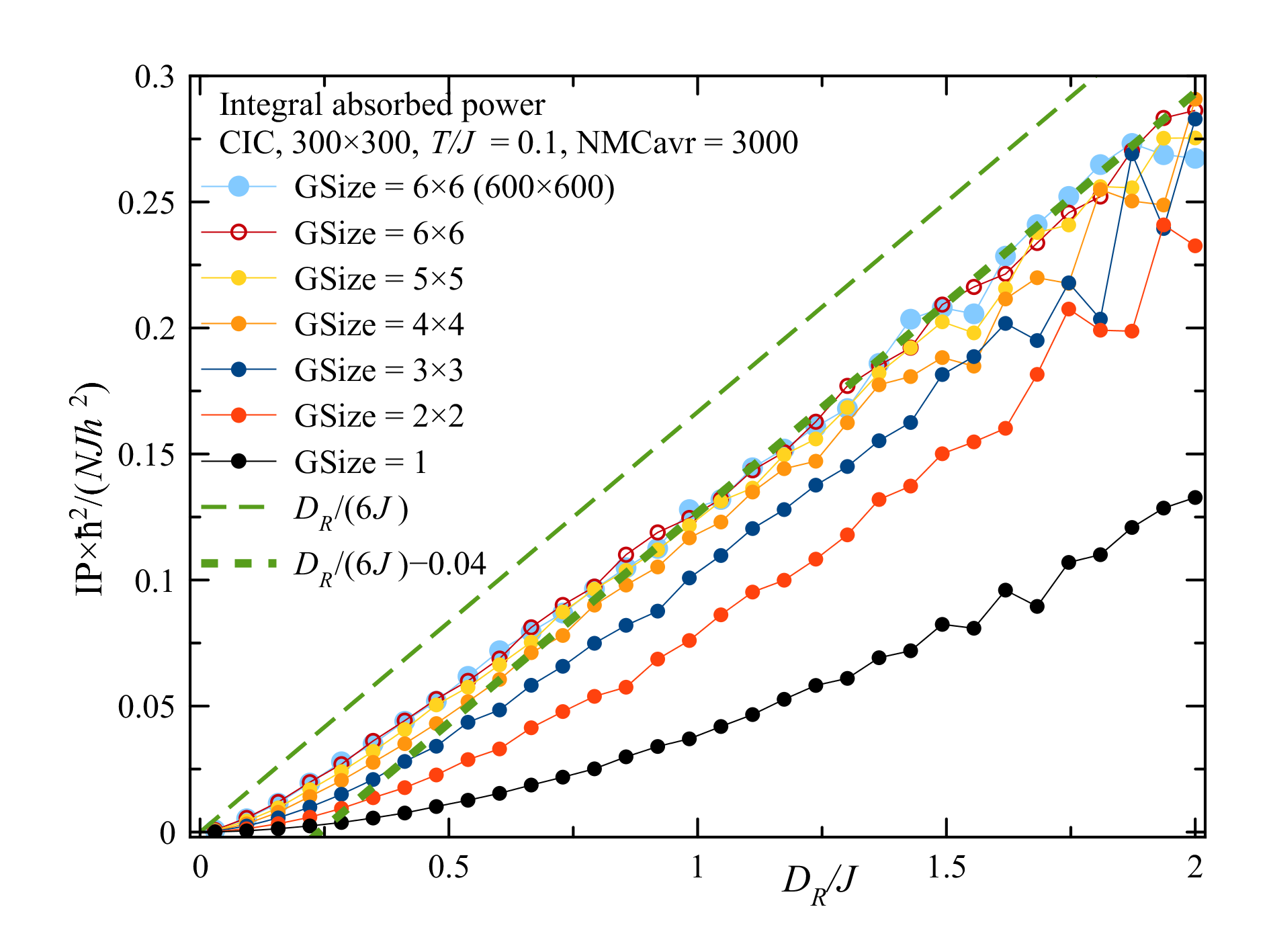}
\caption{Integral power absorption by RA magnets with different grain sizes at large RA.}
\label{largeDR} 
\end{figure}

At large $D_R$ the power absorption by even the smallest grains is dominated by the magnetic anisotropy and is insensitive to the exchange. The IP at large $D_R$ is shown in Fig.\  \ref{largeDR}. In accordance with the expectation, as $D_R$ increases, the slope of curves for all grain sizes tends toward the slope of the line given by Eq.\ (\ref{IP-distributed}). The negative constant term in the integral power absorption at large $D_R$ must be due to domain-wall regions at the grain boundaries. 

\begin{figure}[h]
\centering{}\includegraphics[width=9cm]{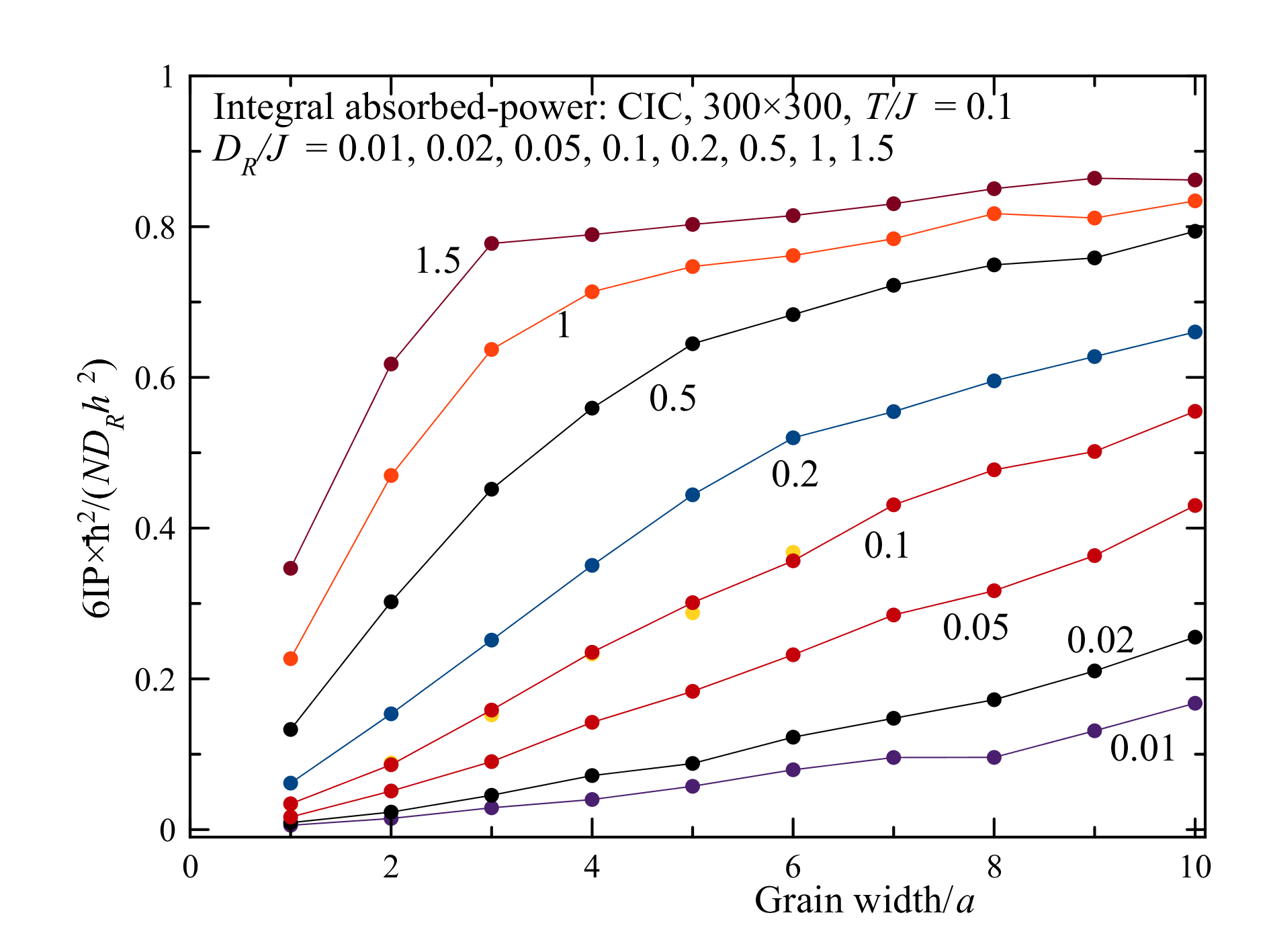}
\caption{Dependence of the integral power absorption on the width of the grain at different strengths of the RA.}
\label{IP-Size} 
\end{figure}

The dependence of the IP on the width of the grain at different $D_R/J$ is shown in Fig. \ref{IP-Size}. One can see a crossover to the case of large independent grains, Eq.\ (\ref{IP-distributed}), on increasing the grain width.

\section{Discussion}
\label{Discussion}

In this paper, we have computed the integral power absorption of electromagnetic radiation (integral over frequency, IP) by random-anisotropy magnets. We assume that the magnet is made of a nonconducting magnetic material or sintered from coated conducting magnetic grains of a size that is small compared to the skin depth of the electromagnetic radiation.

Application of the sum rule with the subsequent usage of analytics or Monte Carlo permits the computation of the integral absorbed power without the knowledge of the dynamical evolution of the system needed for the tour-de-force integration of the frequency-dependent absorbed power. Microwave power absorption by amorphous and sintered magnets strongly depends on the underlying model, in particular, on whether random anisotropy is atomic-scale or correlated. The IP is practically the same in spin states of the RA magnet with high and low magnetization, obtained from the initial state with random and collinear initial orientations of spins, respectively. It gradually decreases with temperature.

We studied the disorder at the atomic scale for $D_R \ll J$, when the ferromagnetic correlation length is large compared with the interatomic distance. In this case, the RA dependence of the scaling of the IP on parameters is dominated, up to a logarithm, by a factor $D_R^2/J$. On the contrary, for a conventional ferromagnet with a coherent anisotropy strength $D = D_R$, or a magnet composed of large ferromagnetic grains, the IP of the ferromagnetic resonance would scale as $D_R$. 

Consequently, for an amorphous magnet that is fully disordered at the atomic scale, the proportionality of the IP to $D_R^2/J$ at $D_R \ll J$ reduces the power absorption by a factor $D_R/J$ as compared to the absorption by large weekly interacting randomly oriented ferromagnetic grains. Within our model, the latter, however, is peaked at a single FMR frequency while the absorption by the RA magnet would be broadband.

Variations in the shape of large grains and the resulting variation in the surface magnetic anisotropy would result in the finite absorption bandwidth. That bandwidth, however, can hardly compete with the bandwidth in an amorphous ferromagnet or a magnet sintered of nanograins where it is determined by a much stronger effect of exchange coupling of the grains with different orientations of the anisotropy axes that creates domain walls at grain boundaries.

Our analysis of that problem reveals that the scaling of the IP on parameters for grains of size $R_a$  is dominated by $(D_R^2/J) (R_a/a)^2$. This answer is valid up to $R_a/a \sim (J/D_R)^{1/2}$. At such a grain size or such a value of the amorphous structure factor, the IP approaches that of a conventional ferromagnet but the absorption remains broadband. 

This determines the optimal amorphous structure factor or the optimal grain size, $(J/D_R)^{1/2}a$, for which both the absorption and the bandwidth are maximal. Notice that this size coincides with the domain wall thickness in a conventional ferromagnet. Typically is would be smaller than the skin depth for the microwaves. 

We hope that our findings will provide guidance for manufacturing magnetic materials with strong broadband absorption of electromagnetic radiation. 
\\
\section*{ACKNOWLEDGEMENT}

This work has been supported by Grant No. FA9550-20-1-0299 funded by the Air Force Office of Scientific Research.

\end{document}